\documentclass[12pt]{article}
\usepackage{amsmath,amssymb} 
\usepackage{subcaption}   

\usepackage{amsfonts}
\usepackage{enumerate}
\usepackage{ytableau}
\usepackage{hyperref}
\numberwithin{equation}{section}
\usepackage{cite}
\usepackage{bm}
\usepackage{tikz}
\usepackage{blkarray}  
\begin{document}

\def\im{\text{i}}
\def\eqa{\begin{eqnarray}}
\def\eqae{\end{eqnarray}}
\def\be{\begin{equation}}
\def\ee{\end{equation}}
\def\bea{\begin{eqnarray}}
\def\eea{\end{eqnarray}}
\def\ba{\begin{array}}
\def\ea{\end{array}}
\def\bd{\begin{displaymath}}
\def\ed{\end{displaymath}}
\def\eg{{\it e.g.~}}
\def\ie{{\it i.e.~}}
\def\Tr{{\rm Tr}}
\def\tr{{\rm tr}}
\def\>{\rangle}
\def\<{\langle}
\def\a{\alpha}
\def\b{\beta}
\def\c{\chi}
\def\del{\delta}
\def\e{\epsilon}
\def\f{\phi}
\def\vf{\varphi}
\def\tvf{\tilde{\varphi}}
\def\g{\gamma}
\def\h{\eta}
\def\j{\psi}
\def\k{\kappa}
\def\l{\lambda}
\def\m{\mu}
\def\n{\nu}
\def\w{\omega}
\def\p{\pi}
\def\q{\theta}
\def\r{\rho}
\def\s{\sigma}
\def\t{\tau}
\def\u{\upsilon}
\def\x{\xi}
\def\z{\zeta}
\def\D{\Delta}
\def\F{\Phi}
\def\G{\Gamma}
\def\J{\Psi}
\def\L{\Lambda}
\def\W{\Omega}
\def\P{\Pi}
\def\Q{\Theta}
\def\S{\Sigma}
\def\U{\Upsilon}
\def\X{\Xi}
\def\nab{\nabla}
\def\pa{\partial}
\newcommand{\lra}{\leftrightarrow}
\def\co{\mathcal{O}}
\def\cl{\mathcal{L}}
\newcommand{\bc}{{\mathbb{C}}}
\newcommand{\br}{{\mathbb{R}}}
\newcommand{\bz}{{\mathbb{Z}}}
\newcommand{\bp}{{\mathbb{P}}}

\def\({\left(}
\def\){\right)}
\def\nn{\nonumber \\}
\def\Poincare{Poincar\'e }
\newcommand{\red}{\textcolor[RGB]{255,0,0}}
\newcommand{\blue}{\textcolor[RGB]{0,0,255}}
\newcommand{\green}{\textcolor[RGB]{0,255,0}}
\newcommand{\cyan}{\textcolor[RGB]{0,255,255}}
\newcommand{\magenta}{\textcolor[RGB]{255,0,255}}
\newcommand{\yellow}{\textcolor[RGB]{255,255,0}}
\newcommand{\sky}{\textcolor[RGB]{135, 206, 235}}
\newcommand{\orange}{\textcolor[RGB]{255, 127, 0}}
\def\d{\operatorname{d}}

\title{
  \begin{center}
    \textbf{The Holography of Spread Complexity} \\
    \textit{A Story of Observers}
  \end{center}
}
\vspace{14mm}
\author{Zhehan Li$^1$\footnote{lizhehan@sxu.edu.cn} and Jia Tian$^{1,2}$
\footnote{wukongjiaozi@ucas.ac.cn}
}
\date{}
\maketitle

\begin{center}
	{\it $^1$State Key Laboratory of Quantum Optics and Quantum Optics Devices, Institute of Theoretical Physics, Shanxi University, Taiyuan 030006, P.~R.~China\\
 \vspace{2mm}
		$^2$Kavli Institute for Theoretical Sciences (KITS),\\
		University of Chinese Academy of Science, 100190 Beijing, P.~R.~China 
	}
\vspace{10mm}
\end{center}

\makeatletter
\def\blfootnote{\xdef\@thefnmark{}\@footnotetext}  
\makeatother

\begin{abstract}
Building on the pioneering work of \cite{Caputa:2024sux}, we propose a holographic description of spread complexity and its rate in 2D CFTs. By exploiting 
$SL(2,\mathbb{R})$ symmetry, we explicitly construct the Krylov basis, expressing spread complexity as a linear combination of generator expectation values. Within the AdS/CFT correspondence, we translate these boundary expectations directly into bulk kinematic variables. These findings suggest that spread complexity manifests as the energy measured by a bulk observer, with its rate corresponding to the radial momentum.
\end{abstract}

\baselineskip 18pt
\newpage

\tableofcontents

\section{Introduction}
Quantum complexity, originating from quantum information theory, quantifies the minimal size of a quantum circuit required to prepare a target state from a reference state. In recent years, seminal developments \cite{Chapman:2021jbh,Baiguera:2025dkc} have established its pivotal role in the AdS/CFT correspondence \cite{Maldacena:1997re,Witten:1998qj,Gubser:1998bc}, offering profound insights into how spacetime geometry emerges from underlying quantum information.

As first conjectured in \cite{Susskind:2014rva}, the universal linear growth of black hole interiors after thermalization \cite{Hartman:2013qma} should be dual to the growth of quantum complexity. Several holographic proposals have been formulated to capture this behavior \cite{Susskind:2014rva,Brown:2015bva,Couch:2016exn}. However, a subsequent proliferation of holographic proposals revealed an infinite class of gravitational observables sharing these properties \cite{Belin:2021bga,Belin:2022xmt}. The multiplicity of holographic complexity is not unexpected, as the definition of quantum complexity itself carries a fundamental ambiguity stemming from the choice of reference state and gate set. This ambiguity is conceptually analogous to that encountered in relativity, where different observers may obtain different measurement outcomes. On the other hand, due to this inherent ambiguity and the computational challenges involved, establishing a precise quantitative comparison between quantum complexity and its holographic duals remains a significant challenge.

A promising alternative is provided by Krylov complexity \cite{Parker:2018yvk}, recently reviewed in \cite{Nandy:2024htc}, which characterizes the growth of operator size during time evolution. Regarded as a new paradigm for complexity \cite{Baiguera:2025dkc}, it is closely related to spread complexity \cite{Balasubramanian:2022tpr}, which measures the dispersion of an initial quantum state in the Hilbert space. Although the precise relation between Krylov/spread complexity and circuit complexity remains an open question, both exhibit the expected behaviors of complexity in chaotic systems \cite{Balasubramanian:2022tpr,Erdmenger:2023wjg}. Crucially, they offer a tractable framework for quantifying quantum unitary evolution. If these quantities admit holographic descriptions, direct comparisons become feasible. Indeed, in lower-dimensional models such as the DSSYK model \cite{Berkooz:2018jqr,Berkooz:2018qkz,Garcia-Garcia:2017pzl}, it has been demonstrated \cite{Lin:2022rbf,Rabinovici:2023yex,Jian:2020qpp,Ambrosini:2024sre,Aguilar-Gutierrez:2025pqp} that spread complexity matches the renormalized length of the wormhole connecting two boundaries exactly in JT gravity \cite{Jackiw:1984je,Teitelboim:1983ux}.

Prior to the advent of Krylov and spread complexity, a notable conjecture related the operator size (complexity) $C_{\mathcal{O}}$ to the momentum $p_{\text{infalling}}$ of a particle falling into the bulk \cite{Susskind:2014jwa,Susskind:2018tei,Susskind:2019ddc,Brown:2018kvn,Susskind:2020gnl,Magan:2018nmu,Barbon:2020uux,Barbon:2020olv,Barbon:2019tuq,Lin:2019kpf,Lin:2019qwu,Ageev:2018msv,Ageev:2018nye}:
\be
\frac{d C_{\mathcal{O}}}{dt}\propto p_{\text{infalling}}.
\ee
On general grounds, both sides of this correspondence are ambiguous: the definition of operator complexity is non-unique, and the specific momentum to be compared is ill-defined. Despite these subtleties, this relation has been verified \cite{Lin:2019qwu} within the JT/SYK duality \cite{Kitaev:2015}, building on earlier studies \cite{Qi:2018bje,Roberts:2018mnp} of operator growth in the SYK model \cite{Sachdev:1992fk}. Equipped with the formalism of spread complexity, this relation has recently been revisited in holographic CFTs \cite{Caputa:2024sux} (see also \cite{Fan:2024iop,He:2024pox, Aguilar-Gutierrez:2025kmw}). In particular, it was proposed that the spread complexity rate is proportional to the so-called ``proper momentum.'' However, the justification for selecting this specific momentum, which relies on a preferred coordinate system, remains mysterious.

In this work, we establish a systematic holographic framework for spread complexity in 2D CFTs, improving upon the mapping approach of \cite{Caputa:2024sux}. By exploiting $SL(2,\mathbb{R})$ symmetry, we explicitly construct the Krylov basis without iterative algorithms. Utilizing the AdS/CFT extrapolated dictionary, and under well-defined assumptions, we directly translate boundary operator expectations into bulk kinematic quantities. Consequently, we refine and prove the conjecture posed in \cite{Caputa:2024sux}. Furthermore, this translation suggests a holographic interpretation wherein spread complexity corresponds to the energy measured by a bulk observer, with its rate corresponding to the measured radial momentum. Crucially, this ``measured momentum'' is invariant under radial coordinate redefinitions, providing a covariant explanation for the proper momentum proposal in \cite{Caputa:2024sux}. Within this prototypical setting, the problem of quantum complexity ambiguities can be partially resolved: according to the AdS/CFT correspondence, the ambiguity in the field theory's initial state maps to a corresponding ambiguity in the gravity theory, while the ambiguity in the gate set relates to a stabilizer group transformation.

The remainder of this paper is organized as follows. In Section 2, we present our proposal within global AdS$_3$, explicitly constructing the Krylov basis and establishing the geometric interpretation of spread complexity as a state distance. In Section 3 and 4, we extend our analysis to Poincar\'e and Rindler AdS$_3$ spacetimes, demonstrating the robustness of our observer-dependent framework across different coordinate systems. Finally, in Section 5, we comment on extensions to higher dimensions and discuss open issues in the Conclusion.

\section{The proposal}
In this section, we establish our proposal within the context of global AdS$_3$ spacetime. The central objectives are threefold: to demonstrate that spread complexity can be computed via a geometric method, to derive its holographic description directly from the AdS/CFT dictionary, and to interpret its growth rate as a measured quantity associated with a specific bulk observer. We summarize the core elements of this framework at the end of this section.

\subsection{The Set-up}
We consider global AdS$_3$ spacetime with the metric
\be
ds^2=-\cosh^2\rho dt^2+d\rho^2+\sinh^2\rho d\phi^2,\quad \phi\in[0,2\pi),\label{globalmetric}
\ee
whose asymptotic boundary is a cylinder. To study the CFT on the cylinder, it is convenient to perform a Wick rotation $t\to -\im t_E$ and introduce the coordinates
\be
z=e^{t_E+\im \phi},\quad \bar{z}=e^{t_E-\im \phi},
\ee
such that states in the CFT on the time slice $t_E=0$ are prepared by a Euclidean path integral on the unit disk: $|z|\leq 1$. The generators of 2D conformal symmetry are given by
\be
L_n=\oint_{|z|=1} \frac{dz}{2\pi \im} z^{n+1} T(z),\quad \bar{L}_n=\oint_{|z|=1} \frac{d\bar{z}}{2\pi \im }\bar{z}^{n+1}\bar{T}(\bar{z}),
\ee
where $T(z)$ and $\bar{T}(\bar{z})$ are the holomorphic and anti-holomorphic components of the energy-momentum tensor. In the absence of operator insertions, the prepared state is the vacuum state $|\Omega\rangle$. Excited states are obtained by inserting local operators $\co(z_i,\bar{z}_i)$ within the unit disk into the path integral. Through this operator-state correspondence, the behavior of operator growth under time evolution can be captured by the spread complexity of the corresponding state. For simplicity, we focus on a scalar primary operator of conformal dimension $\Delta$; thus, the state can be expressed as
\be
|\psi(t)\rangle=\frac{1}{\mathcal{N}}e^{-\im H t}\co(z,\bar{z})|\Omega\rangle,
\ee
where $\mathcal{N}$ is a normalization factor. According to the extrapolated dictionary of AdS/CFT, the initial state is dual to a bulk state of the form
\be
\lim_{\rho\to \infty} \hat{\phi}(\rho,t,\phi)|\Omega\rangle_{\text{AdS}},
\ee
where $\hat{\phi}$ is the dual local scalar field operator. Therefore, near the AdS boundary, the CFT state is dual to a local scalar field state. Moreover, in the semi-classical regime where $\Delta \gg 1$ but $\Delta\ll \frac{1}{G_N}$, the scalar field can be approximated by a probe free massive particle. 

Next, we determine the initial position of the particle dual to the initial state $|\psi_0\rangle=\co(z_0)|\Omega\rangle$. In the bulk, this state is also prepared by a Euclidean path integral from $t_E=-\infty$ to $t_E=0$. Since we neglect the backreaction of the particle, the spacetime remains (Euclidean) global AdS$_3$. In the semi-classical regime, the path integral of the particle can be approximated by the contribution of the classical saddle point, namely the geodesic. To complete the spacetime from $t_E=-\infty$ to $t_E=\infty$, we consider the density matrix $\rho=|\psi_0\rangle\langle \psi_0|=\co|\Omega\rangle\langle \Omega|\co^\dagger$. Therefore, the dual gravity solution of the density matrix $|\psi_0\rangle\langle \psi_0|$ is the Euclidean global AdS$_3$ spacetime with a massive particle propagating from the boundary point $z=z_0$ to another point $z_0^*=1/z_0$. By exploiting spacetime isometry, we can always set $z_0$ to be real without loss of generality.

\begin{figure}[hbt]
\begin{centering}
\includegraphics[scale=0.65]{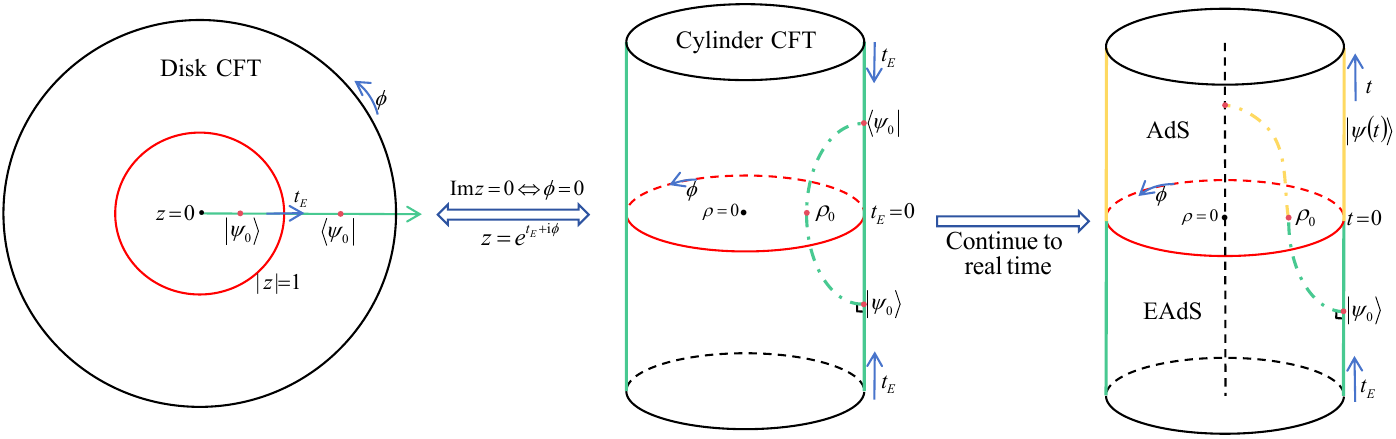}
\caption{Depiction of the holographic construction of the quantum state $|\psi(t)\rangle$.}\label{statef}
\end{centering}
\end{figure}

The geodesic connecting these two points in Euclidean global AdS$_3$ is given by
\be
\tanh\rho=\tanh\rho_0\cosh t_E,
\ee
with
\be
z_0=\tanh\frac{\rho_0}{2}. \label{relation}
\ee
Therefore, the state $|\psi_0\rangle$ is dual to a massive particle located at position $\rho=\rho_0$ on the $t=0$ Cauchy slice. To obtain the time-evolved state $|\psi(t)\rangle$, we replace the upper half of Euclidean AdS$_3$, which is dual to $\langle \psi_0|$, with the corresponding Lorentzian spacetime. The entire construction is illustrated in Figure \ref{statef}.
In conclusion, the state $|\psi(t)\rangle$ is dual to the geodesic trajectory
\be
\tanh\rho=\tanh\rho_0\cos t,\quad t\geq 0. \label{globalgeodesic}
\ee
Note that using the operation
\be
e^{-\rho\frac{L_{1}-L_{-1}+\bar{L}_{1}-\bar{L}_{-1}}{2}}\co(z_0)|\Omega\rangle,
\ee
we can move the inserted operator to another possible location $\tilde{z}_0$ given by
\be
\tilde{z}_0=\frac{z_0+\tanh\frac{\rho}{2}}{1+z_0\tanh\frac{\rho}{2}}=\tanh(\rho_0+z_0),\quad \tilde{\bar{z}}_0=\frac{\bar{z}_0+\tanh\frac{\rho}{2}}{1+\bar{z}_0\tanh\frac{\rho}{2}}=\tanh(\rho_0+\bar{z}_0).
\ee
This implies that the operator $\frac{L_{1}-L_{-1}+\bar{L}_{1}-\bar{L}_{-1}}{2}$ generates translations along the $\rho$ direction, which coincides with the radial momentum operator introduced in \cite{Guo:2022syo}. Actually, we can introduce the $SL(2,\mathbb{R})$ group with generators
\bea
&&l_0=L_0+\bar{L}_0,\quad l_{+1}=L_{1}+\bar{L}_1,\quad l_{-1}=L_{-1}+\bar{L}_{-1},\\
&&[l_n,l_m]=(n-m)l_{n+m}\, ,
\eea
which governs the dynamics of the particle or the CFT state. For example, starting from the trivial geodesic described by $\rho=0$, we can obtain other non-trivial geodesics by applying these $SL(2,\mathbb{R})$ transformations, except for those generated by $l_0$. Even though the duality between the state \eqref{mainstate} and the geodesic \eqref{globalgeodesic} is strictly valid when $\rho\to\infty \, (z_0\to 1)$, we assume that in the semi-classical regime it holds for a generic choice of $z_0$. 
Therefore, we have the following one-to-one mappings:
\bea
&&\co(z_0)|\Omega\rangle  \,  \leftrightarrow \, \text{geodesic} \, \leftrightarrow \, \text{coset } SL(2,\mathbb{R})/U(1),
\eea
Thus, we can keep track of the ambiguities arising from different choices of initial states. Since the particle exhibits no motion along the $\phi$ direction, its dynamics are effectively confined to an AdS$_2$ subspace, thereby reducing the symmetry of the system to $SL(2,\mathbb{R})$, as defined above. Under the $U(1)$ stabilizer group transformation, the initial remains invariant, whereas $SL(2,\mathbb{R})$ generators undergo a rotation. Specifically, for the initial state $\co(z=0)|\Omega\rangle$, where the stabilizer group is generated by $l_0$, the transformation yields:
\be 
e^{\im \epsilon l_0}l_\pm e^{-\im \epsilon l_0}\to e^{\pm \im \epsilon}l_{\pm}.
\ee 
Physically, this can be understood as a change of quantum gates, even though, as we shall demonstrate, it does not change the spread complexity. 
\subsection{The Spread Complexity}
In global AdS$_3$ spacetime, the Hamiltonian $H_{\text{CFT}}$ generating time translation is $L_0+\bar{L}_0=l_0$, so the state corresponding to $\co(z(t))$ is
\be
|\psi(t)\rangle=e^{-\im l_0 t}|\psi_0\rangle=e^{-\im l_0 t}\co(z)|\Omega\rangle,\quad |\psi_0\rangle\equiv\co(z)|\Omega\rangle. \label{mainstate}
\ee
Following the Lanczos algorithm, the Krylov basis is obtained by Gram-Schmidt orthonormalization of the states $\{|\psi_0\rangle, H_{\text{CFT}}|\psi_0\rangle, H^2_{\text{CFT}}|\psi_0\rangle\}$. In the Krylov basis $\{|K_n\rangle\}$, the Hamiltonian takes the form of a tridiagonal matrix, and the quantum state can be expanded as
\be
|\psi(t)\rangle=\sum_{k=0}^\infty \psi_n(t)|K_n\rangle.
\ee
The Krylov complexity, or more precisely, the spread complexity, is defined as the expectation value of the number operator
\be
C(t)=\langle \psi(t)|\hat{n}|\psi(t)\rangle=\sum_{n=0}^\infty n |\psi_n(t)|^2.
\ee
This procedure is difficult to implement, particularly in field theory where the inner product may not be well-defined. Typically, Krylov complexity is obtained from the Lanczos coefficients (the matrix elements of the Hamiltonian), which are determined iteratively \cite{Viswanath1994}. For the particular state \eqref{mainstate} corresponding to a local operator, \cite{Caputa:2024sux} obtained the spread complexity by mapping the time evolution to an effective $SL(2,\mathbb{R})$ dynamics via the identification of return amplitudes. With this approach, the only input is the CFT two-point function, and all information about the Krylov basis remains implicit. However, as noted, this information is crucial for deriving the holographic dual. We therefore adopt a method for constructing the Krylov basis that does not employ the Lanczos algorithm or iterative procedures.

For the primary operator $\co(z)$ with conformal dimension $\Delta=h+\bar{h}$, when it is inserted at the origin of the unit disk, the prepared state is called the asymptotic state
\be
\co(0)|\Omega\rangle=|\Delta\rangle,
\ee
which satisfies
\be
l_0|\Delta\rangle=\Delta|\Delta\rangle,\quad l_{+1}|\Delta\rangle=0. \label{eigen}
\ee
In other words, the state $|\Delta\rangle$ is the highest weight state of the $\mathfrak{sl}_2$ algebra spanned by $\{l_0,l_{\pm 1}\}$. It generates the representation $\{|\Delta,n\rangle\}$ of the $\mathfrak{sl}_2$ algebra:
\begin{align}
&|\Delta,n\rangle\equiv\sqrt{\frac{\Gamma(2\Delta)}{n!\Gamma(2\Delta+n)}}l_{-1}^n|\Delta\rangle,\quad \langle \Delta,n|\Delta,m\rangle=\delta_{n,m},\\
&l_0|\Delta,n\rangle=(\Delta+n)|\Delta,n\rangle,\\
&l_{+1}|\Delta,n\rangle=\sqrt{n(2\Delta+n-1)}|\Delta,n-1\rangle,\\
&l_{-1}|\Delta,n\rangle=\sqrt{(n+1)(2\Delta+n)}|\Delta,n+1\rangle.
\end{align}
Since $|\Delta\rangle$ is an eigenstate of the Hamiltonian $l_{0}$, the state remains unchanged under time evolution, yielding vanishing spread complexity.

When the operator is inserted at a generic point, the corresponding state becomes a linear combination of states $|\Delta,n\rangle$:
\begin{equation}
|\psi_0\rangle=\mathcal{O}(z)|\Omega\rangle=e^{z l_{-1}}\mathcal{O}(0)|\Omega\rangle=\sum_{n=0}^\infty \frac{z^n}{n!}l_{-1}^n|\Delta\rangle=\sum_{n=0}^\infty \sqrt{\frac{\Gamma(2\Delta+n)}{n!\Gamma(2\Delta)}} z^n|\Delta,n\rangle.
\end{equation}
As this state is no longer an eigenstate of $l_0$, one can construct the Krylov subspace $\{|\psi_0\rangle,l_0|\psi_0\rangle,l_0^2|\psi_0\rangle,\ldots\}$. However, this approach is cumbersome as the state $|\psi_0\rangle$ is expressed in an inconvenient basis. To find a more efficient representation, we exploit the fact that $|\psi_0\rangle$ can be viewed as a highest weight state of the $\mathfrak{sl}_2$ algebra spanned by a different set of generators $\{\mathcal{L}_0,\mathcal{L}_{\pm 1}\}$ (see e.g. \cite{Chattopadhyay:2023fob}). This can be verified by starting from \eqref{eigen} and applying a generic unitary $SL(2,\mathbb{R})$ transformation:
\begin{align}
&e^{\mathrm{i} Q}l_0 e^{-\mathrm{i} Q}e^{\mathrm{i} Q}\mathcal{O}(0)|\Omega\rangle=\Delta e^{\mathrm{i} Q}\mathcal{O}(0)|\Omega\rangle \quad\Rightarrow\quad \mathcal{L}_0\mathcal{O}(z)|\Omega\rangle=\Delta\mathcal{O}(z)|\Omega\rangle,\\
&Q=\alpha l_0+\frac{\beta-\mathrm{i} \gamma}{2}l_{+1}+\frac{\beta+\mathrm{i} \gamma}{2}l_{-1},\quad \alpha,\beta,\gamma \in \mathbb{R}, \label{chargeQ}
\end{align}
where
\begin{equation}
\mathcal{O}(z) \propto e^{\mathrm{i} Q}\mathcal{O}(0)e^{-\mathrm{i} Q},\quad \mathcal{L}_n=e^{\mathrm{i} Q}l_n e^{-\mathrm{i} Q}.
\end{equation}
The explicit relations between $l_n$ and $\cl_n$ are:
\begin{align}
&z=\frac{B}{D}, \quad |z|\leq 1,\label{conz}\\
&\mathcal{L}_0=-AB l_{-1}+(AD+BC)l_{0}-CD l_{+1},\label{t1}\\
&\mathcal{L}_{+1}=B^2 l_{-1}-2BDl_0+D^2l_{+1},\label{t2}\\
&\mathcal{L}_{-1}=A^2l_{-1}-2ACl_{0}+C^2 l_{+1},\label{t3}
\end{align}
with
\begin{equation}
\begin{pmatrix}
A&B\\
C&D
\end{pmatrix}=\begin{pmatrix}
\cosh\left(\frac{\sqrt{\delta}}{2}\right)+\frac{\mathrm{i}\alpha\sinh\left(\frac{\sqrt{\delta}}{2}\right)}{\sqrt{\delta}} & \frac{\mathrm{i}(\beta+\mathrm{i}\gamma)\sinh\left(\frac{\sqrt{\delta}}{2}\right)}{\sqrt{\delta}} \\
\frac{-(\gamma+\mathrm{i}\beta)\sinh\left(\frac{\sqrt{\delta}}{2}\right)}{\sqrt{\delta}} & \cosh\left(\frac{\sqrt{\delta}}{2}\right)-\frac{\mathrm{i}\alpha\sinh\left(\frac{\sqrt{\delta}}{2}\right)}{\sqrt{\delta}}
\end{pmatrix}, \label{ABCD}
\end{equation}
where $\delta=\beta^2+\gamma^2-\alpha^2$. Inverting relations \eqref{t1}--\eqref{t3}, we obtain
\begin{equation}
l_0=(AD+BC)\mathcal{L}_{0}+BD\mathcal{L}_{-1}+AC\mathcal{L}_{+1}.
\end{equation}
Therefore, the state \eqref{mainstate} becomes
\begin{equation}
|\psi(t)\rangle=e^{-\mathrm{i}[(AD+BC)\mathcal{L}_{0}+BD\mathcal{L}_{-1}+AC\mathcal{L}_{+1}]t}|\Delta\rangle_{z},\quad |\Delta\rangle_{z}\equiv \mathcal{O}(z)|\Omega\rangle=|\psi_0\rangle. \label{ext}
\end{equation}
The Krylov basis can now be directly identified as
\begin{equation}
|K_n\rangle=|\Delta,n\rangle_z,\quad |\Delta,n\rangle_z\equiv \sqrt{\frac{\Gamma(2\Delta)}{n!\Gamma(2\Delta+n)}}\mathcal{L}_{-1}^n|\Delta\rangle_z,
\end{equation}
and the resulting spread complexity is given by
\begin{equation}
C(t)=\frac{\langle \psi(t)|\mathcal{L}_0|\psi(t)\rangle}{\langle \psi(t)|\psi(t)\rangle}-\Delta. \label{rsc}
\end{equation}
A few comments are in order. The $\mathfrak{sl}_2$ generators $\mathcal{L}_n$ with $n=0,\pm 1$ can be extended to the full set of Virasoro algebra generators, and the state $|\Delta\rangle_z$ is promoted to a primary state of the Virasoro algebra. When $z$ is real, this Virasoro algebra arises from the so-called $\text{M\"obius}$ quantization of 2D CFT studied in \cite{Okunishi:2016zat}. When the operator is inserted at the boundary of the unit disk, i.e., $z=1$, which corresponds to the $t=0$ time slice in the 2D cylinder spacetime, the prepared state is ill-defined. A commonly used regularization scheme is to perform an infinitesimal Euclidean time evolution \cite{Nozaki:2014hna,Caputa:2014vaa}:
\begin{equation}
\mathcal{O}(t=0)|\Omega\rangle  \to e^{-\epsilon l_0}\mathcal{O}(t=0)|\Omega\rangle.
\end{equation}
This is the type of state considered in \cite{Caputa:2024sux}. The physical interpretation of this infinitesimal Euclidean time evolution is now clear: it simply moves the operator into the unit disk along the radial direction, and the regulator $\epsilon$ is related to the initial position of the massive particle through the relation \eqref{relation} with $z_0=e^{-\epsilon}$. The spread complexity given in \eqref{rsc} is invariant under the $SL(2,\mathbb{R})$ unitary transformation. By substituting \eqref{t1}, we can rewrite it as
\begin{equation}
C(t)=(AD+BC) \frac{\langle \psi(t)|l_0|\psi(t)\rangle}{\langle \psi(t)|\psi(t)\rangle}-AB\frac{\langle \psi(t)|l_{-1}|\psi(t)\rangle}{\langle \psi(t)|\psi(t)\rangle}-CD \frac{\langle \psi(t)|l_{+1}|\psi(t)\rangle}{\langle \psi(t)|\psi(t)\rangle}-\Delta. \label{mainK}
\end{equation}
This is one of the main results of this section, showing that the spread complexity depends not only on the state $|\psi(t)\rangle$ but also on the information about the Krylov basis (the representation) encoded in the unitary transformation generated by $Q$ \eqref{chargeQ}, or equivalently, the matrix \eqref{ABCD}.

\subsection{The Spread Complexity as a State Distance}
Note that there are three real parameters $(\alpha,\beta,\gamma)$ in the charge $Q$ \eqref{chargeQ}, but only one complex constraint \eqref{conz}. Consequently, one free parameter remains in the expression \eqref{ext}, and correspondingly, the resulting spread complexity \eqref{rsc} also contains a free parameter. For a fixed operator location at $z$, a residual $U(1)$ group transformation exists within the global $SL(2,\mathbb{R})$, known as the little group or stabilizer group. The stabilizer group is parameterized as
\be
e^{\im \tilde{Q}}=e^{\im (\tilde{\alpha} l_0+\frac{\tilde{\beta}-\im \tilde{\gamma}}{2}l_{+1}+\frac{\tilde{\beta}+\im \tilde{\gamma}}{2}l_{-1})} \label{stable}
\ee
and generates the specific $SL(2,\mathbb{R})$ transformation
\be
e^{\im \tilde{Q}}\co(z)e^{-\im \tilde{Q}}\propto\co(\tilde{z}),\quad \tilde{z}=\frac{\tilde{A}z+\tilde{B}}{\tilde{C}z+\tilde{D}}.
\ee
The parameters are subjected to the condition
\be
\tilde{z}=\frac{\tilde{A}z+\tilde{B}}{\tilde{C}z+\tilde{D}}=z,
\ee
where the relationships between $\tilde{A},\tilde{B},\tilde{C},\tilde{D}$ and $\tilde{\alpha},\tilde{\beta},\tilde{\gamma}$ are also given in \eqref{ABCD}.
The solution to this condition is indeed one-dimensional:
\be
\tilde{\alpha}=\frac{\im (z^2-1)\tilde{\gamma}-(z^2+1)\tilde{\beta}}{2z},\quad \text{Im}(z)\tilde{\beta}=\text{Re}(z)\tilde{\gamma}.
\ee
The action of the stabilizer group is to change one suitable representation to another, or equivalently, to change one set of the $\mathfrak{sl}_2$ generators $\{\cl_0,\cl_{\pm 1}\}$ to another set $\{\tilde{\cl}_0,\tilde{\cl}_{\pm 1}\}$, which mimics the change of quantum gate sets in the context of quantum complexity, albeit it does not change the value of the complexity.

To generalize and better understand the spread complexity defined above, we introduce the notion of \textbf{Krylov state distance}. We have shown that any scalar primary operator $\co(z_1)$ defines a highest weight state $|\Delta\rangle_{z_1}$ and the representation $\{|\Delta,n\rangle_{z_1}\}$. The same scalar primary operator at another location $\co(z_2)$ can be written as a linear combination
\bea
&&|\co(z_2)\rangle=\mathcal{N}e^{\im Q}|\co(z_1)\rangle=\sum_{n=0}^\infty c_n |\Delta,n\rangle_{z_1},\\
&&Q=\alpha l_0+\frac{\beta-\im \gamma}{2}l_{+1}+\frac{\beta+\im \gamma}{2}l_{-1},\quad \alpha,\beta,\gamma \in \mathbb{R},
\eea
subject to
\be
\frac{A z_1+B}{C z_1+D}=z_2.
\ee
Mimicking the spread complexity, we define the Krylov state distance as
\be
C_{z_1\to z_2}=\sum_{n=0}^\infty |c_n|^2 n. \label{c12n}
\ee
Recall that spread complexity characterizes operator growth under time evolution. Similarly, the Krylov state distance can be understood as describing operator growth under a specific $SL(2,\mathbb{R})$ flow generated by $Q$. It is worth noting that Krylov complexity in 2D CFTs subjected to deformed $SL(2,\mathbb{R})$ Hamiltonians has been extensively studied in \cite{Malvimat:2024vhr}.

\begin{figure}[hbt]
\begin{centering}
\includegraphics[scale=0.75]{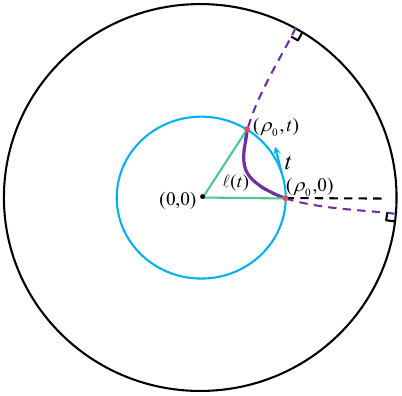}
\caption{The circular blue line denotes the time evolution, the purple line, which is perpendicular to the hyperbolic boundary denoted by the black circle, is the hyperbolic geodesic connecting the starting and ending points.}\label{timetr}
\end{centering}
\end{figure}

It is clear that the Krylov state distance is invariant under the $SL(2,\mathbb{R})$ transformation following
\bea
&&e^{\im \Q'}|\co(z_2)\rangle=\mathcal{N}e^{\im Q'}e^{\im Q}|\co(z_1)\rangle,\to |\co(z_2')\rangle=\mathcal{N}' e^{\im\mathcal{Q}}|\co(z_1')\rangle,\\
&&C_{z_1\to z_2}=C_{z_{1}'\to z_{2}'} \label{uni}
\eea
where
\bea
z'_{1,2}=\frac{A' z_{1,2}+B'}{C'z_{1,2}+D'},\quad \mathcal{Q}=\alpha l'_0+\frac{\beta-\im \gamma}{2}l'_{+1}+\frac{\beta+\im \gamma}{2}l'_{-1},\quad l'_n\equiv e^{\im Q}l_n e^{-\im Q}.
\eea
Without loss of generality, we can transform $z_1$ to the origin, leave the other point as a general point $z_2=\tanh\frac{\rho}{2}e^{\im \phi}$ and write
\be
\co(z_2)|\Omega\rangle=e^{-\frac{\rho}{2}(l_{+1}e^{-\im \phi}-l_{-1}e^{\im \phi})}|\co(0)\rangle. \label{repstate}
\ee
Then, we find that the Krylov state distance is
\be
C_{z_1\to z_2}=\Delta \frac{2|z_{21}|^2}{1-|z_{21}|^2}=\Delta(\cosh\rho-1)=2\Delta \sinh ^2(\frac{\rho}{2} ) \label{d12}
\ee
which coincides with the isometric invariant of the hyperbolic disk, where the variable $\rho$ represents the hyperbolic distance (geodesic distance in the hyperbolic space). This is not a coincidence because the invariance of the distance complexity under $SL(2,\mathbb{R})$ transformation implies that it must be a function of the hyperbolic distance, since $SL(2,\mathbb{R})$ is the isometry group of the hyperbolic space, whose metric is
\be
ds^2_{\text{hyperbolic}}=\frac{4dz d\bar{z}}{(1-|z|^2)^2}=d\rho^2+\sinh^2\rho d\phi^2. \label{disk}
\ee
With this geometric interpretation, we can immediately write down the Krylov state distance between two generic points:
\be
C_{z_1\to z_2}=\Delta \frac{2|z_{12}|^2}{(1-|z_1|^2)(1-|z_2|^2)}=\Delta(\cosh\ell_{12}-1), \label{distance}
\ee
where $\ell_{12}$ is the hyperbolic distance. Furthermore, the representative \eqref{repstate} can also be viewed as an isomorphism from the coset $SL(2,\mathbb{R})/U(1)$ to the unit disk. It is worth mentioning that the state \eqref{repstate} is known as the generalized coherent state \cite{Perelomov:1986uhd}, which is associated with an information geometry equipped with a Fubini-Study metric that corresponds exactly to the hyperbolic disk \eqref{disk}. It turns out that our notion of Krylov state distance is closely related to the proposal in \cite{Chapman:2017rqy}, where state complexity is identified with the geodesic length in the Fubini-Study metric, and in \cite{Lv:2023jbv}, where spread complexity is identified with an embedding coordinate difference in the Hyperbolic disk. Furthermore, in \cite{Caputa:2021sib}, it was proposed that the volume enclosed by a circle in hyperbolic space is proportional to the Krylov complexity of a specific state, up to an overall constant. Other properties, such as the entanglement structure and holographic description of Virasoro coherent states, have been investigated in \cite{Caputa:2022zsr, Liska:2022vrd}. It is also worth mentioning that the state distance \eqref{distance} has a more transparent geometric meaning in terms of the embedding coordinates for the hyperbolic disk:
\be
X_0=\cosh\rho,\quad X_1=\sinh\rho \cos u,\quad X_2=\sinh\rho \sin u,\quad \vec{X}\cdot \vec{X}\equiv X_0^2-X_1^2-X_2^2=1,
\ee
which is equal to the angle (up to a constant shift) between the two points:
\be
C_{\vec{X}\to \vec{Y}}=\Delta (\vec{X}\cdot \vec{Y}-1).
\ee
The spread complexity defined in \eqref{rsc} is equal to the following Krylov state distance:
\be
|\psi(t)\rangle=e^{-\im l_0 t}\co(z_0)|\Omega\rangle=\co(z_0e^{-\im t})|\Omega\rangle,\quad \to \quad C(t)=C_{z_0\to z_0 e^{-\im t}}. \label{timestate}
\ee
The state \eqref{timestate} implies that the time evolution is projected to a circular trajectory in the coset space $SL(2,\mathbb{R})/U(1)$, as shown in Figure \ref{timetr},
and substituting into \eqref{distance}, we can get the corresponding spread complexity
\bea
C(t)=C_{z\to z e^{-\im t}}&=&\Delta (\cosh \ell(t)-1)=\Delta (\cosh^2\rho_0-\sinh^2\rho_0\cos t-1),\\
&=&2\Delta \sinh^2\rho_0 \sin^2\frac{t}{2},\quad z\equiv\tanh\frac{\rho_0}{2}, \label{relation0}
\eea
where we have used the hyperbolic law of cosines. Alternatively, one can obtain the same result from \eqref{mainK} by substituting
\bea
&&\begin{pmatrix}
A&B\\
C&D
\end{pmatrix}=\left(
\begin{array}{cc}
\cosh \left(\frac{\rho_0 }{2}\right) &  \sinh \left(\frac{\rho_0 }{2}\right) \\
\sinh \left(\frac{\rho_0  }{2}\right) & \cosh \left(\frac{\rho_0  }{2}\right) \\
\end{array}
\right),\label{thetaABCD}\\
&&\frac{\langle \psi(t)|l_0|\psi(t)\rangle}{\langle \psi(t)|\psi(t)\rangle}=\frac{\langle \psi_0|l_{0}|\psi_0\rangle}{\langle \psi_0|\psi_0\rangle}=\Delta \cosh\rho_0,\label{ev0}\\
&&\frac{\langle \psi(t)|l_{\pm 1}|\psi(t)\rangle}{\langle \psi(t)|\psi(t)\rangle}=e^{\mp \im t}\frac{\langle \psi_0|l_{\pm 1}|\psi_0\rangle}{\langle \psi_0|\psi_0\rangle}=e^{\mp \im t}\Delta \sinh\rho_0,\label{evpn}
\eea
where we have used the identities
\be
e^{\im l_0 t}l_{\pm1} e^{-\im l_0 t}=e^{\mp \im t}l_{\pm1}. \label{timeconju}
\ee
This result for spread complexity \eqref{relation0} matches the one derived in \cite{Caputa:2024sux} from the mapping method. Note that our proposal provides a simple geometric approach for computing Krylov state distance under any $SL(2,\mathbb{R})$ flows.

Note that we have identified the hyperbolic disk \eqref{disk} appearing as the manifold of the states $\{\mathcal{O}(z)|\Omega\rangle\}$ with the reduced metric of global AdS$_3$ at the $t=0$ slice
\be
ds_{t=0}^2=d\rho^2+\sinh^2\rho d\phi^2.
\ee
This identification suggests that the Krylov state distance is indeed related to a physical spacetime distance via \eqref{distance}. It should be noted that a similar correspondence between timelike geodesics in AdS and quantum circuits (states) has also been proposed in \cite{Chagnet:2021uvi}.

\subsection{The Holographic Description}
We have shown that the spread complexity is given by \eqref{mainK}, which takes the form of a linear combination of the expectation values of the $SL(2,\mathbb{R})$ generators. Therefore, using the AdS/CFT dictionary to translate these boundary expectation values to bulk ones, we can directly obtain the holographic description.

The conformal transformations generated by these generators can be lifted to isometric transformations in the AdS spacetime, associated with the Killing vector fields $\hat{L}_n (\hat{\bar{L}}_n)$ corresponding to the generators $L_n (\bar{L}_n)$ \cite{Maldacena:1998bw, Balasubramanian:1998sn}. These Killing fields as operators act on the bulk scalar field as\footnote{We have used the same notations $L_n$ for the symmetry generator operators in the CFT and the bulk theory.}
\be
[L_n,\hat{\phi}(0)]=\im \hat{L}_n^\mu \partial_\mu\hat{\phi}
\ee
thus the expectation values can be written as
\be
\langle \Omega|\hat{\phi}(0)e^{\im H t}  L_ne^{-\im H t}\hat\phi(0)|\Omega\rangle_{\text{AdS}}=\im \hat{L}_n^\mu(t) \langle\phi|\partial_\mu |\phi\rangle_{\text{AdS}}\equiv\im \hat{L}_n^\mu(t) \hat{p}_\mu,\quad \hat{L}_n(t)\equiv \widehat{e^{\im H t}  L_ne^{-\im H t}}. \label{bulkexpect}
\ee
In the semi-classical regime, ${p}_\mu\equiv \langle\phi|\partial_\mu |\phi\rangle_{\text{AdS}}$ should reduce to the momentum of the massive particle. In global AdS$_3$ spacetime, described by the metric \eqref{globalmetric}, the Killing vectors associated with $l_n$ are explicitly given by
\bea
&&\hat{l}_0=\im \partial_t,\\
&&\hat{l}_{\pm 1}=\im e^{\pm \im t}\(\cos\phi\tanh\rho \partial_t\pm\im \sin\phi\coth\rho \partial_\phi\mp\im  \cos\phi\partial_\rho\),
\eea
and the tangent vector of the geodesic \eqref{globalgeodesic} is
\be
\hat{u}_\phi=0,\quad \hat{u}_t=-\cosh\rho_0,\quad \hat{u}_\rho=-\sin t\sinh\rho_0. \label{tanv}
\ee
Indeed, as proposed in \eqref{bulkexpect}, we find the following relations:
\bea
\langle \psi(t)|l_0|\psi(t)\rangle&=&\Delta\cosh\rho_0=\im \hat{l}_0^\alpha p_\alpha\Big|_{\text{geodesic}},
\nonumber\\
\langle \psi(t)|l_{\pm}|\psi(t)\rangle&=&e^{\mp \im t}\Delta \sinh\rho_0=\im e^{\mp \im t}\hat{l}_\pm ^\alpha p_\alpha\Big|_{\text{geodesic}}, \label{relations}
\eea
where we have introduced the canonical momentum of the particle $p_\alpha=\Delta \hat{u}_\alpha$ and used the identity \eqref{timeconju}. The relations \eqref{relations} suggest that expectation values of the generators $l_n$ can be interpreted as measured energy and momentum by an observer using a set of (unorthonormal) tetrads $\{\im \hat{l}_0,\im e^{ \mp \im t}\hat{l}_\pm \}$. Substituting \eqref{thetaABCD}-\eqref{evpn} into \eqref{mainK}, we find
\be
\cl_0=\cosh\rho_0 l_0-\sinh\rho_0\frac{l_{+1}+l_{-1}}{2},\label{cl0}
\ee
such that the spread complexity can be written compactly as
\be
C(t)+\Delta=\hat{e}_0^\alpha(\rho) p_\alpha, \label{re1}
\ee
where
\bea
\hat{e}_0(\rho)&=&\im \cosh\rho_0 \hat{l}_0-\im \sinh\rho_0\frac{e^{-\im t}\hat{l}_{+1}+e^{\im t}\hat{l}_{-1}}{2}
\nonumber\\
&=&(\cos\phi\sinh\rho_0\tanh \rho-\cosh \rho_0)\partial_t,
\eea
which means that the spread complexity is proportional to the energy of the particle measured by an observer moving along the timelike vector $\hat{e}_0$. Similarly, except for $\hat{e}_0$ we can associate the observer with two other local Lorentz orthonormal frames (tetrad):
\bea
\hat{e}_{\pm1 } &=&\im\frac{\sinh\rho_0}{2}(\frac{e^{\mp\im t}\hat{l}_{\pm1}}{\tanh\frac{\rho_0}{2}}-2\hat{l}_0+\tanh\frac{\rho_0}{2}e^{\pm \im t}\hat{l}_{\mp1})
\nonumber\\
&=&(\sinh\rho_0-\cos\phi \cosh\rho_0 \tanh\rho)\partial_t \mp \im \sin\phi\cosh \rho \partial_\phi\pm\im \cos\phi\partial_\rho
\eea
The components of the measured momentum are then:
\begin{align}
P_0 &= \hat{e}_0 \cdot \hat{p}\Big|_{\text{geodesic}}  = \Delta (\cosh^2\rho_0 - \sinh^2\rho_0 \cos t), \\
P_\pm &= \hat{e}_\pm \cdot \hat{p}\Big|_{\text{geodesic}} = \Delta \sinh\rho_0\(\cosh\rho_0(\cos(t)-1)\mp \im \sin t\) .
\label{measuredr}
\end{align}
Next, we derive the correspondence proposed in \cite{Caputa:2024sux}, which states that the spread complexity rate is proportional to the (proper) radial momentum:
\be
\frac{dC(t)}{dt}\propto p_\rho.
\ee
Starting from \eqref{rsc} and \eqref{cl0}, we can derive the complexity rate
\be
\frac{dC(t)}{dt}=\im \frac{\langle\psi(t)|[l_0,\cl_0]|\psi(t)\rangle}{\langle\psi(t)|\psi(t)\rangle}=\sinh\rho_0 \frac{\langle\psi(t)|\frac{\im}{2}(l_{+1}-l_{-1})|\psi(t)\rangle}{\langle\psi(t)|\psi(t)\rangle}.
\ee
Following our earlier argument, the operator $\frac{\im}{2}(l_+-l_-)$ corresponds to the vector field
\be
\frac{\im}{2}(\hat{e}_+-\hat{e}_-)=\sin\phi \coth\rho \partial_\phi-\cos\phi \partial_\rho.
\ee
Therefore, we immediately find that the complexity rate is proportional to the measured radial momentum:
\be
\frac{d C(t)}{dt}=\sinh\rho_0\frac{\im}{2}(\hat{e}_+-\hat{e}_-)^\alpha p_\alpha\Big|_{\text{geodesic}}=-\sinh\rho_0p_\rho, \label{rate1}
\ee
which happens to be proportional to the radial momentum in the coordinates \eqref{globalmetric}. It should be pointed out that there is a subtlety in the correspondence between the complexity rate and radial momentum. The radial momentum itself is not invariant under redefinitions of the radial coordinate, which raises the question: which radial coordinate should be used? In \cite{Caputa:2024sux} (see also \cite{Fan:2024iop, He:2024pox}), the radial coordinate is chosen such that the radial momentum coincides with the complexity rate. In contrast, our framework not only provides an explicit construction of the holographic description but also provides an invariant quantity---the measured radial momentum---that is independent of the choice of radial coordinate. This and the proposed correspondence \eqref{bulkexpect} constitute the main results of our work.

\begin{figure}[hbt]
\begin{centering}
\includegraphics[scale=0.9]{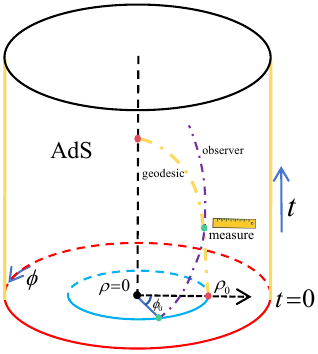}
\caption{Illustration of our proposal: A specific observer interacts with the dual particle and performs a measurement. The spread complexity corresponds to the measured energy, while the complexity rate is identified with the radial momentum observed by that observer.}\label{momentum}
\end{centering}
\end{figure}

Before turning to additional examples, we summarize the core elements of our proposal:
\begin{itemize}
\item The AdS/CFT extrapolated dictionary facilitates the construction of bulk local fields from boundary local CFT operators.
\item Near the AdS boundary, the bulk local field reduces to a single local CFT operator, establishing a one-to-one mapping between the bulk state $\hat{\phi}|\Omega\rangle_{\text{AdS}}$ and a boundary state $\co|\Omega\rangle_{\text{CFT}}$. In the semi-classical limit, the bulk scalar field acts as a probe particle, with its initial position linked to the CFT operator's location via the Euclidean geodesic equation.
\item For a scalar primary boundary operator $\co$, the states $\co(z)|\Omega\rangle_{\text{CFT}}$ represent generalized coherent states, forming a metric coset manifold.
\item Since both are invariant quantities, the spread complexity (or Krylov state distance) is related to the geodesic distance on this coset manifold. Moreover, the Krylov state distance admits a representation as a linear combination of symmetry generator expectation values.
\item Via the AdS/CFT correspondence, these expectation values can also be computed in the bulk theory through \eqref{bulkexpect}, taking the form of projected momenta. We interpret these as the momenta measured by an observer utilizing the tetrad frame constructed from the Killing fields associated with the symmetry generators. This geometric interpretation naturally extends to both the spread complexity and its rate of growth. Our proposal can be illustrated with Figure \ref{momentum}.
\end{itemize}
In the following sections, we demonstrate how to apply our proposal to obtain results in other AdS spacetimes straightforwardly.
\section{The \Poincare AdS$_3$}
The metric of the \Poincare AdS$_3$ spacetime is given by
\be
ds^2=\frac{-dT^2+dX^2+dZ^2}{Z^2},
\ee
which is related to the global AdS$_3$ metric \eqref{globalmetric} via the following coordinate transformations:
\bea
Z&=&\frac{1}{\sinh (\rho ) \cos (\phi )+\cosh (\rho ) \cos (t)},
\nonumber\\
X&=&Z \sinh\rho \sin\phi,\label{LTr}\\
T&=& Z\cosh\rho \sin t.
\eea
To study the dual CFT, we perform a Wick rotation of the time coordinate $T=-\im T_E$ and introduce holomorphic and anti-holomorphic coordinates:
\be
\xi=T_E+\im X,\quad \bar{\xi }=T_E-\im X.
\ee
The coordinate transformations \eqref{LTr} imply the conformal map between the global CFT and the \Poincare CFT:
\be
z=\frac{1+\xi }{1-\xi },\quad \xi=\frac{z-1}{z+1} \label{GTPRelation},
\ee
under which the hyperbolic plane becomes the \Poincare half-plane with the metric
\be
ds^2=4\frac{d\xi d\bar{\xi }}{(\xi+\bar{\xi })^2}=\frac{dT_E^2+dX^2}{T_E^2}.
\ee
The length of the geodesic connecting arbitrary points $P_1=(T_{E,1},X_1)$ and $P_2=(T_{E,2},X_2)$ is given by
\be
\ell_{12}=2\text{arcsinh}\frac{|P_1-P_2|}{2\sqrt{T_{E,1}T_{E,2}}}.
\ee
Using the relation \eqref{distance}, we immediately obtain the Krylov state distance in \Poincare CFT:
\be
C_{(T_{E,1},X_1)\to (T_{E,2},X_2)}=\Delta(\cosh\ell_{12}-1)=\Delta\frac{(T_{E,1}-T_{E,2})^2+(X_1-X_2)^2}{2 T_{E,1}T_{E,2}}.
\ee
The states in the \Poincare CFT are obtained via a Euclidean path integral from $T_E=-\infty$ to $T_E=0$, which corresponds to the unit disk in the $z$ coordinates. Without loss of generality, suppose the operator is initially inserted at $\xi=\xi_0$, preparing the state:
\be
|\psi_0\rangle=\co(\xi_0=-T_{E,0})|\Omega\rangle,\quad T_{E,0}>0. \label{pinitial}
\ee
The Lorentzian time evolution is $\xi=\xi_0+\im T$, so the spread complexity and its rate become:
\be
C(T)=C_{\xi_0,\xi_0+\im T}=\Delta\frac{T^2}{2 T_{E,0}^2},\quad \frac{dC(T)}{dT}=\Delta\frac{T}{T_{E,0}},
\ee
which exactly matches the result derived in \cite{Caputa:2024sux}. It is important to note that this result cannot be obtained simply by applying the coordinate transformation \eqref{LTr} to \eqref{relation0}. The reason lies in the fact that the Hamiltonian in the \Poincare spacetime is generated by $H_{\text{\Poincare}}=L_{-1}^P+\bar{L}_{-1}^P\equiv l_{-1}^P$, where the $SL(2,\mathbb{R})$ generators are defined as
\be
L_n^P=\int_{-\im \infty}^{\im \infty}\frac{d\xi }{2\pi \im}\xi^{n+1}T(\xi ),\quad  \bar{L}_n^P=\int_{-\im \infty}^{\im \infty}\frac{d\bar{\xi} }{2\pi \im}\bar{\xi}^{n+1}\bar{T}(\bar{\xi} ),
\ee
and their relation to the global $SL(2,\mathbb{R})$ generators can be derived by substituting into the conformal transformation \eqref{GTPRelation}:
\be
L_n^P=\oint dz \frac{(z+1)^2}{2}\left(\frac{z-1}{z+1}\right)^{n+1}T(z),\quad \bar{L}_n^P=\oint d\bar{z} \frac{(\bar{z}+1)^2}{2}\left(\frac{\bar{z}-1}{\bar{z}+1}\right)^{n+1}\bar{T}(\bar{z}).
\ee
Explicitly, the relations between the two sets of generators $l_n=L_n+\bar{L}_n$ and $l_n^P=L_n^P+\bar{L}_n^P$ are:
\bea
&&l_0^P=\frac{l_{+1}-l_{-1}}{2},\quad l_{\pm 1}^P=\mp l_0+\frac{l_{+1}+l_{-1}}{2}, \label{relation2}\\
&&l_0=-\frac{l_{+1}^P-l_{-1}^P}{2},\quad l_{\pm 1}=\pm l_0^P+\frac{l_{+1}^P+l_{-1}^P}{2}.
\eea
The conjugate of the initial state \eqref{pinitial} is
\be
\langle \psi_0|=\langle \Omega|\co(\xi=T_{E,0}),
\ee
and as previously discussed, the density matrix $|\psi_0\rangle \langle \psi_0|$ is dual to the Euclidean \Poincare spacetime with a massive particle propagating along the Euclidean geodesic
\be
Z^2+T_E^2=T_{E,0}^2\, .
\ee
Furthermore, the time-evolved state $|\psi(t)\rangle=e^{\im L_{-1}^P t}|\psi_0\rangle$ is dual to the Lorentzian geodesic
\be
Z^2-T^2=T_{E,0}^2,\quad T\geq 0,
\ee
with tangent vector $\hat{u}=(u^T,u^X,u^Z)$ and momentum components:
\bea
&&u^T=\frac{T^2+T_{E,0}^2}{T_{E,0}},\quad u^X=0,\quad u^Z=\frac{T\sqrt{T^2+T_{E,0}^2}}{T_{E,0}},\\
&&p_T=-\frac{\Delta}{T_{E,0}},\quad p_X=0,\quad p_Z=\frac{\Delta T}{T_{E,0}\sqrt{T^2+T_{E,0}^2}}.
\eea
The generators $l_n^P$ correspond to the following (normalized) Killing vectors:
\bea
&&\hat{l}_0^P=-\left(Z\partial_Z+T\partial_T+X\partial_X\right),\\
&&\hat{l}_{-1}^P=\im \partial_T,\\
&&\hat{l}_{+1}^P=-\im \left(2TZ\partial_Z+\left(T^2+X^2+Z^2 \right)\partial_T+2TX\partial_X\right).
\eea
Similarly, we can verify the relations \eqref{bulkexpect}:
\bea
&&\langle \psi(t)|l^P_0|\psi(t)\rangle =\langle \psi_0|( -\im T)l^P_{-1}+l_0^P|\psi_0\rangle=-\Delta\frac{\im T}{T_{E,0}}=\im(-\im T \hat{l}_{-1}^{P,\alpha}+\hat{l}^{P,\alpha}_0)p_\alpha\Big|_{\text{geodesic}},
\nonumber\\
&&\langle \psi(t)|l^P_{-1}|\psi(t)\rangle=\langle \psi_0|l_{-1}^P|\psi_0\rangle=\frac{\Delta}{T_{E,0}}=\im l_{-1}^{P,\alpha}p_{\alpha}\Big|_{\text{geodesic}},\\
&&\langle \psi(t)|l_{+1}^P|\psi(t)\rangle=\langle \psi_0|(-\im T)^2l_{-1}^P-2\im T l_0^P+l_{+1}^P|\psi_0\rangle=-\Delta\frac{T^2+T_{E,0}^2}{T_{E,0}}
\nonumber\\
&&\qquad \qquad \qquad ~ =\im \left((-\im T)^2\hat{l}_{-1}^{P,\alpha}-2\im T\hat{l}_0^{P,\alpha}+\hat{l}_{+1}^{P,\alpha} \right)p_\alpha\Big|_{\text{geodesic}}.
\eea
We find it more convenient to compute the expectation values of the generators in the original $z$ coordinates. For example,
\be
{\langle \psi(t)|l^P_{-1}|\psi(t)\rangle}={\langle \psi_0|l^P_{-1}|\psi_0\rangle}=\langle \co(z_0)\left(l_0+\frac{l_{+1}+l_{-1}}{2}\right)\co(z_0)\rangle=\frac{\Delta}{T_{E,0}},
\ee
where $z_0=\frac{1-T_{E,0}}{1+T_{E,0}}$. In the last section, we have shown that for the state $|\psi_0\rangle=\co(z_0=\tanh\frac{\rho_0}{2})|\Omega\rangle$, the spread complexity is determined by the generator
\be
\cl_0=\cosh\rho_0 l_0-\frac{1}{2}\sinh\rho_0(l_{+1}+l_{-1}).
\ee
Transforming into the \Poincare coordinates, we obtain
\be
\cl_0^P=-\frac{1}{2}\left(\frac{l_{+1}^P}{{T_{E,0}}}-{T_{E,0}}l^P_{-1}\right)
\ee
such that the spread complexity can be written as
\bea
&&C(T)=\hat{e}_0^\alpha p_\alpha\Big|_{\text{geodesic}}-\Delta=\frac{T^2}{2T_{E,0}^2},\\
&&\hat{e}_0=-\frac{\im}{2}\left(\frac{(-\im T)^2\hat{l}_{-1}^{P}-2\im T\hat{l}_0^{P}+\hat{l}_{+1}^{P}}{T_{E,0}}-T_{E,0}\hat{l}_{-1}^P\right)
\nonumber\\
&&\quad=\frac{-1}{T_{E,0}}\left(TZ\partial_Z+(2T^2+X^2+Z^2-T_{E,0}^2)\partial_T+TX\partial_X\right),
\eea
The spread complexity growth rate is then given by
\be
\frac{dC(T)}{dT}=\im {\langle\psi(t)|[l_{-1}^P,\cl_0^P]|\psi(t)\rangle}=\frac{\im}{T_{E,0}} {\langle\psi(t)|l_0^P|\psi(t)\rangle}=\frac{(\im T \hat{l}_{-1}^{P,\alpha}-\hat{l}^{P,\alpha}_0)p_\alpha\Big|_{\text{geodesic}}}{T_{E,0}}.
\ee
Since the operator $\im T \hat{l}^P_{-1}-\hat{l}_0^P$ corresponds to the vector field
\be
\im T \hat{l}^P_{-1}-\hat{l}_0^P=Z\partial_Z+X\partial_X\equiv \hat{e}_Z,
\ee
we find that the complexity rate is proportional to the measured radial momentum:
\be
\frac{dC(T)}{dT}=\frac{\hat{e}^\alpha_Z p_\alpha\Big|_{\text{geodesic}}}{T_{E,0}}=\frac{Z}{T_{E,0}}p_Z,
\ee
rather than the canonical momentum $p_Z$. We emphasize again that the fact $\hat{e}^\alpha_Z p_\alpha \neq p_Z$ simply reflects the coordinate choice. In terms of the proper radial coordinate the \Poincare AdS metric becomes
\be
ds^2=d\rho^2+e^{2\rho}(-dT^2+dX^2),
\ee
the radial vector becomes $\hat{e}_\rho=X\partial_X-\partial_\rho$
with the measured radial momentum equal to the (opposite of) proper radial momentum:
\be
P_\rho=\hat{e}_\rho^\mu p_{\mu}\Big|_{\text{geodesic}}=-p_\rho,
\ee
as proposed in \cite{Caputa:2024sux}. Therefore, our proposal explains why the proper radial momentum proposal gives the correct result for the complexity growth rate.
\section{The Rindler AdS$_3$}
The Rindler spacetime (also known as the decompactified BTZ black hole) consists of four distinct wedges. Since our arguments rely primarily on the isometries of the spacetime, we expect that the results obtained in the Rindler spacetime will carry over directly to the (single-sided) BTZ black hole. We will insert the operator in the right wedge, where the metric is given by
\begin{align}
ds^2 &= -\sinh^2\rho_r dt_r^2 + d\rho_r^2 + \cosh^2\rho_r d\phi_r^2, \\
\rho_r &> 0,\quad -\infty < t_r < \infty,\quad -\infty < \phi_r < \infty.
\end{align}
The left wedge can be obtained from the right one via the analytic continuation $t_l=-t_r+\im \pi,\, \phi_l=-\phi_r$. The Rindler space is dual to the thermofield double state $\Psi_{\text{TFD}}$, which is invariant under the combined $SL(2,\mathbb{R})$ transformations:
\be
L_n^R|\Psi_{\text{TFD}}\rangle=(L_n^r-(-1)^n L_{-n}^l)|\Psi_{\text{TFD}}\rangle=0.
\ee
Since the operator is inserted only in the right wedge, its dynamics in the left wedge are trivial. Therefore, we only need to restrict our attention to the right wedge. The coordinate transformations between global AdS and the right wedge of Rindler space are:
\bea
&&\cosh\rho\cos t=\cosh\rho_r\cosh\phi_r,\quad \cosh\rho \sin t=\sinh\rho_r\sinh t_r,\\
&&\sinh\rho\sin \phi=\cosh\rho_r\sinh\phi_r,\quad \sinh\rho \cos \phi=\sinh\rho_r\cosh t_r.
\eea
Similarly, we define an Euclidean coordinate on the boundary of the Rindler spacetime as \cite{Goto:2017olq}
\be
\zeta=e^{\phi_r-\im \tau_r},
\ee
where $\tau_{r,l}=\im t_{r,l}$. These coordinate transformations imply the conformal map
\be
z=-\im\frac{1+\im\zeta}{1-\im \zeta},\quad \zeta=\im \frac{1-\im z}{1+\im z}.
\ee
They also yield the following relations between the $SL(2,\mathbb{R})$ generators \cite{Goto:2017olq}:
\bea
&&L_0=-\im \frac{L_{+1}^R+L_{-1}^R}{2},\quad L_{\pm 1}=-\im L_0^R\pm \frac{L_{+1}^R-L_{-1}^R}{2},
\nonumber\\
&&L_0^R=\im \frac{L_{+1}+L_{-1}}{2},\quad L_{\pm 1}^R=\im L_0\pm \frac{L_{+1}-L_{-1}}{2},
\nonumber\\
&&\bar{L}_0=-\im \frac{\bar{L}_{+1}^R+\bar{L}_{-1}^R}{2},\quad \bar{L}_{\pm 1}=\im \bar{L}_0^R\pm \frac{\bar{L}_{+1}^R-\bar{L}_{-1}^R}{2},
\nonumber\\
&&\bar{L}_0^R=-\im \frac{\bar{L}_{+1}+\bar{L}_{-1}}{2},\quad \bar{L}_{\pm 1}^R=-\im \bar{L}_0\pm \frac{\bar{L}_{+1}-\bar{L}_{-1}}{2}.\label{RTG}
\eea
Note that $L_n^R(\bar{L}_n^R)$ are anti-Hermitian operators, and the Rindler Hamiltonian is given by $H^{\text{Rindler}}=-\im(L_0^R-\bar{L}_0^R)$. Therefore, we define the effective $SL(2,\mathbb{R})$ generators as
\be
l_0^R=L_0^R-\bar{L}_0^R,\quad l_\pm^R=L_\pm^R-\bar{L}_\mp^R.
\ee
In the $\zeta$ coordinates, the hyperbolic plane has the metric
\be
ds^2=-\frac{4d\zeta d\bar{\zeta}}{(\zeta-\bar{\zeta})^2}.
\ee
It is convenient to introduce a new variable
\be
\zeta=e^{\phi_r} \frac{\im+e^{\rho_r}}{-\im+e^{\rho_r}},
\ee
which allows us to rewrite the metric in a more familiar form:
\be
ds^2=d\rho_r^2+\cosh^2\rho_r d\phi_r^2. \label{rm}
\ee
The length of the geodesic connecting arbitrary two points $P_1=(\rho_{r,1},\phi_{r,1})$ and $P_2=(\rho_{r,2},\phi_{r,2})$ is given by
\be
\ell_{12}=\text{arccosh}\left(\cosh \rho_{r,1}\cosh \rho_{r,2}\cosh(\phi_{r,2}-\phi_{r,1})-\sinh \rho_{r,1}\sinh \rho_{r,2}\right)
\ee
and the corresponding state distance becomes
\be
C_{P_1\to P_2}=\Delta(\cosh\ell_{12}-1)=\Delta\left(\left(\cosh \rho_{r,1}\cosh \rho_{r,2}\cosh(\phi_{r,2}-\phi_{r,1})-\sinh \rho_{r,1}\sinh \rho_{r,2}\right)-1\right).
\ee
Under the Lorentzian time evolution $(\rho_{0},\phi_{r,0})\to (\rho_{0},\phi_{r,0}+t_r)$, the spread complexity and its rate become:
\be
C(t_r)=\Delta(\cosh^2\rho_{0} \cosh t_r-\sinh^2\rho_{0}-1),\quad \frac{d C(t_r)}{d t_r}=\Delta\cosh^2\rho_{0} \sinh t_r\, ,
\ee
which agree with the results in \cite{Caputa:2024sux}.
Without loss of generality, we can set $\phi_{r,0}=0$, so the initial state is
\be
|\psi_0\rangle=\co(\theta_{r,0})|\Psi_{\text{TFD}}\rangle.
\ee
Following previous arguments, we can find that time-evolved state $|\psi(t)\rangle$ is dual to the geodesic
\be
\tanh\rho_r\cosh t_r=\tanh\rho_0,\quad \rho_0=\theta_{r,0} ,\quad t_r\geq 0.
\ee
As in the global spacetime, identifying $\rho_0=\theta_{r,0}$ suggests that the hyperbolic space \eqref{rm} can be identified with the $t_r=0$ time slice. The tangent vector along this geodesic and the associated momentum are:
\bea
\hat{u}&=&\sinh\rho_0(\coth^2\rho_0\cosh^2 t_r-1)\partial_{t_r}-\cosh\rho_0\sinh t_r \partial_\rho,\\
p_{t_r}&=&-\Delta\sinh^2\rho_0,\quad p_{\phi_{r}}=0,\quad p_{\rho}=-\Delta\cosh\rho_0\sinh t_r.
\eea
In the right wedge of the Rindler space, the Killing vectors are
\bea
\hat{l}_{0}^R &=&-\partial_{t_r}\\
\hat{l}_{\pm 1}^R &=&-e^{\pm t_r}(\cosh\phi_r \coth\rho_r \partial_{t_r}\pm\sinh \phi_r\tanh\rho_r \partial_{\phi_r}\mp\partial_{\rho_r}),
\eea
and they relate to the expectation values of the corresponding generators as follows:
\bea
{\langle \psi(t_r)|l^R_0|\psi(t_r)\rangle}&=&\Delta \sinh\rho_0=\hat{l}_0^{R,\alpha}p_{\alpha},\\
{\langle \psi(t_r)|l^R_{\pm}|\psi(t_r)\rangle}&=&e^{\mp  t_r}\Delta \cosh\rho_0=e^{\mp  t_r}\hat{l}_\pm^{R,\alpha}p_{\alpha}.
\eea
Note that comparing with \eqref{bulkexpect}, there is a factor of $\im$ difference due to the choice of $H^{\text{Rindler}}=-\im l_0^R$.
Substituting \eqref{RTG} into the expression for \eqref{cl0} and retaining only the right-wedge components, we find that the spread complexity corresponds to the generator
\be
\cl^R_0=\im(\sinh\rho_0 l_0^R-\cosh\rho_0\frac{l^R_{+1}+l^R_{-1}}{2})\, .
\ee
which leads to
\bea
C(t_r)+\Delta &=&\hat{e}_0^\alpha p_\alpha\Big|_{\text{geodesic}}=\Delta\left(\cosh^2\rho_0\cosh t_r-\sinh^2\rho_0\right),\\
\hat{e}_0&=&\im(\sinh\rho_0 \hat{l}_0^R-\cosh\rho_0\frac{e^{-t_r}\hat{l}^R_{+1}+e^{t_r}\hat{l}^R_{-1}}{2})
\nonumber\\
&=&\im(\cosh\rho_0 \cosh\phi_r \coth\rho_r- \sinh\rho_0)\partial_{t_r}.
\eea
Similarly, we find that the complexity rate is given by
\be
\frac{dC(t_r)}{dt_r}={\langle\psi(t)|[l^R_0,\cl^R_0]|\psi(t)\rangle}=\cosh\rho_0{\langle\psi(t)|\im\frac{l_{-1}^R-l_{+1}^R}{2}|\psi(t)\rangle},
\ee
which can be written as
\bea
\frac{dC(t_r)}{dt_r}&=&-\im \cosh\rho_0 \hat{e}_\rho^{\alpha}p_\alpha\Big|_{\text{geodesic}}=\Delta\cosh^2\rho_{0} \sinh t_r,\\
\hat{e}_\rho &\equiv& \frac{\im}{2}\left(e^{t_r}\hat{l}_{-1}-e^{-t_r}\hat{l}_{+1}\right)=-\im(\cosh\phi_r\partial_{\rho_r}-\sinh\phi_r\tanh\rho_r\partial_{\phi_r}).
\eea
The factor $-\im$ still comes from the choice of $H^{\text{Rindler}}=-\im l_0^R$.
\section{Comments on AdS$_{d+1}$}
The extrapolated dictionary and the correspondence between conformal symmetry generators and Killing vectors extend naturally to higher dimensions. In the semi-classical regime, the primary state $\co|\Omega\rangle$ can still be approximated by a free probe particle. Consider the AdS$_{d+1}$ metric
\be
ds^2=-\cosh^2\rho dt^2+d\rho^2+\sinh^2\rho d\vec{x}^2,
\ee
where $\vec{x}^2=1$ parameterizes the sphere $S^{d-1}$. The boundary CFT lives on $\mathbb{R}\times S^{d-1}$. Performing a Wick rotation $t\to -\im \tau$, the conformal symmetry of the CFT is described by the group $SO(1,d+1)$. We denote the generators of $SO(1,d+1)$ as $L_{ab},\, a,b=-1,0,\dots,d$, acting in the embedding space $\mathbb{R}^{1,d+1}$. These relate to the ``physical'' generators in the CFT as follows:
\bea
L_{\mu\nu}=M_{\mu\nu},\quad L_{-1,0}=D,\quad L_{0,\mu}=\frac{1}{2}(P_\mu+K_\mu),\quad L_{-1,\mu}= \frac{1}{2}(P_\mu-K_\mu).
\eea
Assuming at time $t=0$, the probe particle is at rest at the position $\rho=\rho_0, \, \vec{x}=\vec{x}_0=(1,0,0,\dots)$. Since it does not have an initial momentum, the particle will still move in the AdS$_2$ subspace $(t,\rho)$, thus the symmetry of the effective dynamics is still $SL(2,\mathbb{R})$ whose generators can be identified as $\{D,P_{1},K_1\}\equiv \{l_0,l_{-1},l_{+1}\}$. In particular, the geodesic is still given by \eqref{globalgeodesic}. The time-evolved dual state can be written as
\be
|\psi(t)\rangle=e^{-\im D t}\co(z_0\equiv e^{\tau_0},\vec{x}_0)|\Omega\rangle.
\ee
As in the 2D case, the $\co(\tau_0,\vec{x}_0)|\Omega\rangle$ defines a primary state of the generator defined by
\be
\cl_0=e^{\frac{\rho_0}{2} ( K_1-P_1)}De^{-\frac{\rho_0}{2} ( K_1-P_1)},\quad \tanh\frac{\rho_0}{2}=z_0.
\ee
Hence, the structure of the complexity and its holographic description remain analogous to the 2D case.

\section{Conclusions and Outlook}
In this work, we proposed a systematic holographic description of the spread complexity and its growth rate in 2D CFTs. Differing from the approach adopted in the original work \cite{Caputa:2024sux}, we used the AdS/CFT extrapolated dictionary to translate the calculations in the CFT to the ones in the bulk theory, thus directly finding the holographic description. Observing that the spread complexity and its growth rate are expressed as projected momenta, we interpreted them as measured energy or momentum by an observer, which is manifestly coordinate independent. Moreover, by leveraging the underlying symmetries, we explicitly constructed the Krylov basis, and we mapped the freedom in the choices of initial state and set of quantum gates in defining quantum complexity into the ones in choices of initial position of the particle and the corresponding little group transformation of the $SL(2,\mathbb{R})$ symmetry generators. To describe the spread complexity under general $SL(2,\mathbb{R})$ flows, we introduced the notion of Krylov complexity distance, providing a novel geometric interpretation of the spread complexity and connecting it to Nielsen complexity. We also commented on how our proposal can be easily extended to higher dimensions as long as the effective $SL(2,\mathbb{R})$ symmetry shows up. Below, we outline several open issues and potential directions for future research.

First, throughout this work, we have focused on the semi-classical regime where the conformal dimension $\Delta$ of the primary operator is in the range of $\frac{1}{G_N}\gg \Delta\gg 1$. In this limit, the dual state corresponds to a massive particle propagating along a geodesic in an unperturbed AdS spacetime. When $\Delta\ll 1$, the primary operator should instead be dual to a free bulk scalar field, and the time-evolved state $|\psi(t)\rangle$ is described by the wavefunction $\Psi(\rho,t,\phi)=\langle \hat{\phi}(\rho,t,\phi)|\psi(t)\rangle$. In this regime, we expect the conserved charges $\hat{Q}_n$ to be replaced by appropriate bulk expectation values such as $\langle \Psi^*, \hat{L}_n^\mu \partial_\mu\Psi\rangle$, while the spread complexity may still admit an interpretation as the field energy measured by a given observer.

For $\Delta\sim \frac{1}{G_N}$, we can still treat the primary operator as a massive particle, but the backreaction of the particle becomes non-negligible. The resulting geometry develops conical singularities around the worldline of the particle. When the particle is at rest at the center of the global AdS$_3$, the bulk solution corresponds to the conical AdS$_3$ spacetime. For off-center static particles, recent studies suggest that the dual geometry can be described by a three-dimensional C-metric \cite{Astorino:2011mw,Xu:2011vp,Arenas-Henriquez:2022www,Arenas-Henriquez:2023hur,Kubiznak:2024ijq,Fontana:2024odl,Cisterna:2023qhh,Tian:2023ine,Tian:2024mew}. In the case of a moving particle, the dual geometry has been argued to correspond to a locally quenched spacetime \cite{Nozaki:2014hna}. Importantly, all these geometries are related to global AdS$_3$ via coordinate transformations. Therefore, the worldline of the observer can be extended to these more general settings. A potentially related work is \cite{Erdmenger:2022lov}.

Second, the extrapolated dictionary also exists in dS/CFT \cite{Strominger:2001pn,Witten:2001kn,Maldacena:2002vr} and flat/CFT \cite{Barnich:2010eb,Bagchi:2010zz,Fareghbal:2013ifa,Pasterski:2016qvg,Pasterski:2017kqt} correspondence, so our proposal can potentially be generalized to those theories as well.

Third, we can consider the second derivative of the spread complexity. For example, in the global AdS$_3$, we find
\be
\ddot{C}=\frac{\sinh\rho_0}{2} \frac{\langle\psi(t)|(l_{+1}+l_{-1})|\psi(t)\rangle}{\langle\psi(t)|\psi(t)\rangle}=\Delta\cos t\sinh^2\rho_0=-\sinh\rho_0\dot{p}_\rho,\label{force}
\ee
which shows that the acceleration of the spread complexity is also related to that of the particles, suggesting an analog of Ehrenfest's theorem for spread complexity \cite{Erdmenger:2023wjg}.

Fourth, from Eq.~\eqref{force}, we see that the operator $(l_{+1}+l_{-1})/2$ plays the role of the force. This implies that a state evolved under a more general $SL(2,\mathbb{R})$ Hamiltonian $H=\alpha l_0+\beta\frac{l_{+1}+l_{-1}}{2}$ corresponds to a non-free particle. It would be interesting to explore how the spread complexity encodes features of such accelerated motion.

Fifth, the switchback effect is one of the hallmark features of holographic complexity, reflecting the delayed growth of complexity under perturbation. Within the DSSYK model, signatures of the switchback effect in Krylov complexity have been reported \cite{Ambrosini:2024sre}, and the effect has recently been explicitly observed in \cite{Aguilar-Gutierrez:2025mxf}. The Krylov complexity (and its rate) in the DSSYK model with matter fields has also been thoroughly studied in recent works \cite{Aguilar-Gutierrez:2025pqp,Aguilar-Gutierrez:2025hty}, where a relational interpretation of Krylov complexity has also been proposed.
Understanding whether and how this effect manifests in Krylov complexity within holographic CFTs would be an important step toward solidifying its role as a holographic observable. In a simple hyperbolic disk model \cite{Brown:2016wib}, it was found that the switchback effect can arise from a small perpendicular displacement of the state's flow in Krylov space. From the perspective of our proposal, such a displacement may correspond to a perturbation of the dual particle trajectory, potentially induced by interactions or collisions with other particles.

Sixth, while our holographic model, like most models in the literature on holographic complexity, is a bottom-up construction, it would be both interesting and valuable to explore a top-down holographic realization. See recent works \cite{Das:2024tnw, Fatemiabhari:2025poq, Fatemiabhari:2025cyy,Fatemiabhari:2026rob,Zoakos:2026obl} along this line.

Finally, in our analysis of Rindler spacetime, we inserted the operator only in the right wedge. However, inserting operators in both left and right wedges could prepare states dual to entire spacetime geometries, including the black hole interior. In the semi-classical regime, such configurations may lead to states dual to geodesics crossing the horizon, connecting insertion points in opposite wedges. In such cases, the spread complexity may encode information about the black hole interior. This opens up the exciting possibility of using spread complexity as a probe of quantum gravity effects inside black holes. See \cite{Aguilar-Gutierrez:2025kmw} for recent discussions along these lines.

\section*{Acknowledgments}
We are grateful to Chen Bai, Bowen Chen, Cheng Peng, and Yu-Xuan Zhang for their insightful discussions and valuable perspectives. JT would like to especially thank Pawel Caputa for his helpful clarifications and for answering our many questions during his seminar talk at KITS. This work was supported through visiting scholar funding from the Kavli Institute for Theoretical Sciences (KITS) at the University of Chinese Academy of Sciences (UCAS).

\bibliographystyle{unsrturl}
\bibliography{ref}
\end{document}